\documentclass[10pt,journal]{IEEEtran}
\usepackage{mathtools}
\usepackage{amssymb,amsmath}
\usepackage[T1]{fontenc}
\usepackage{ifpdf}
\usepackage{url}
\ifpdf
\usepackage[pdftex]{graphicx}
\graphicspath{{figs/}}
\DeclareGraphicsExtensions{.pdf,.png,.jpg}
\else
\usepackage[dvips]{graphicx}

\graphicspath{{eps/}}
\DeclareGraphicsExtensions{.eps}
\fi
\usepackage{float}
\usepackage[font=footnotesize]{caption}
\usepackage{subcaption}
\usepackage{setspace}
\usepackage{balance}
\usepackage{algorithmic}
\usepackage{algorithm}
\newcommand{\algorithmicinput}{\textbf{Input:}}
\newcommand{\INPUT}{\item[\algorithmicinput]}

\floatname{algorithm}{Procedure}
\usepackage{tikz}
\usetikzlibrary{matrix,arrows,circuits.ee,circuits.ee.IEC,shapes.geometric,shapes.misc}
\newcommand{\iap}{\textit{DREMS}}
\newcommand{\iapfull}{\textbf{D}istributed \textbf{RE}altime \textbf{M}anaged \textbf{S}ystem}
\newcommand{\ALOOP}[1]{\ALC@it\algorithmicloop\ #1%
  \begin{ALC@loop}}
\newcommand{\ENDALOOP}{\end{ALC@loop}\ALC@it\algorithmicendloop}

\usepackage{multirow}

\begin{document}
\title{ DREMS-OS: An Operating System for Managed Distributed Real-time Embedded Systems }
\vspace{-0.1in}
\author{\IEEEauthorblockN{Abhishek Dubey, Gabor Karsai, Aniruddha Gokhale, William Emfinger, Pranav Kumar\\} 
\IEEEauthorblockA{
  ISIS, Dept of EECS, Vanderbilt University, Nashville, TN 37235, USA \\
}
}


\maketitle

\begin{abstract}
Distributed real-time and embedded (DRE) systems executing mixed
criticality task sets are increasingly being deployed in mobile and
embedded cloud computing platforms, including space applications.
These DRE systems must not only operate over a range of temporal and
spatial scales, but also require stringent assurances for secure
interactions between the system's tasks without violating their
individual timing constraints. 
To address these challenges, this paper describes a novel
distributed operating system focusing on the
scheduler design to support the mixed criticality task sets. 
  Empirical results from experiments involving a
case study of a cluster of satellites emulated in a laboratory
testbed validate our claims.
\end{abstract}


\vspace{-0.1in}
\section{Introduction}
\label{sec:intro}

The emerging realm of mobile and embedded cloud computing, which
leverages the progress made in computing and communication
in mobile devices and sensors necessitates a platform for running distributed, 
real-time, and embedded (DRE) systems.  
Ensembles of mobile
devices are being used as a computing resource in space missions as
well: satellite clusters provide a dynamic environment for deploying and 
managing distributed mission applications; see, \emph{e.g.} NASA's Edison Demonstration of SmallSat Networks,
TanDEM-X, PROBA-3, and Prisma from ESA, and DARPA's System F6.

As an example consider a cluster of satellites that execute
software applications distributed across the satellites.  One application
is a safety-critical cluster flight application (CFA) that controls the
satellite's flight and is required to respond to emergency 'scatter' commands.  
Running concurrently with
the CFA, image processing applications (IPA) utilize the satellites' sensors
and consume much of the CPU resources.  IPAs from different vendors may have different
security privileges and so may have controlled access to sensor data. 
Sensitive camera data must be compartmentalized and must not be shared between these 
IPAs, unless explicitly permitted. These applications must also be isolated from each other to prevent 
performance impact or fault propagation between applications due to lifecycle changes.  However, the isolation should not waste CPU resources when applications are dormant  
because, for example, a sensor is active only in certain segments of the satellite's orbit. 
Other  applications should be able to opportunistically use the CPU during these dormant phases.

One technique for implementing strict application isolation  is temporal and spatial
partitioning of processes (see \cite{ARINC-653}).  Spatial separation provides a
hardware-supported, physically separated memory address space for each process.  Temporal partitioning
provides a periodically repeating fixed interval of CPU time that is
exclusively assigned to a group of cooperating tasks. 
Note that strictly partitioned systems are typically configured with a
static schedule; any change in the schedule requires the system to be
rebooted~\cite{ARINC-653}.   

To address these needs, we have developed an architecture called
 \iapfull\ (\iap)~\cite{6899124}.
This paper focuses on the design and implementation of key components of the operating system layer in \iap.
 It describes the design choices and algorithms used in the design of the \iap\ OS scheduler.  The scheduler supports three criticality levels: critical, application, and best effort.  It  supports temporal and spatial partitioning for application-level
  tasks.   Tasks in a partition are scheduled in a work-conserving   manner.  Through a CPU cap mechanism, it also ensures that no task  starves for the CPU. Furthermore, it allows dynamic reconfiguration of the temporal partitions. We  empirically validated the design in the context of a case study: a managed DRE system running on a laboratory testbed.

The outline of this paper is as follows: Section~\ref{sec:related}
presents the related research; 
Section~\ref{sec:task_model} describes the system model and delves
into the details of the scheduler design;
Section~\ref{sec:experiment} empirically evaluates \iap\ OS in the
context of a representative space application; and finally
Section~\ref{sec:conclusions} offers concluding remarks referring to
future work.

\section{Related Research}
\label{sec:related}

Our approach has been inspired by two domains: mixed criticality systems
and partitioning operating systems. 
A mixed criticality
computing system has two or more criticality levels on a single
shared hardware platform, where the distinct levels are motivated by
safety and/or security concerns. For example, an avionics system can
have safety-critical, mission-critical, and non-critical tasks.

In~\cite{Vestal2007}, Vestal argued that the criticality levels directly
impact the task parameters, especially the worst-case execution time
(WCET). In his framework, each task has a maximum criticality level
and a non-increasing WCET for successively decreasing criticality
levels. For criticality levels higher than the task maximum, the
task is excluded from the analyzed set of tasks. Thus increasing
criticality levels result in a more conservative verification
process.  He extended the response-time analysis
of fixed priority scheduling to mixed criticality task sets.
His  results were later improved by Baruah et
al.~\cite{BaruahRTA4MCS} where an implementation was proposed for
fixed priority single processor scheduling of mixed-criticality tasks
with optimal priority assignment and response-time analysis.

Partitioning operating systems have been applied to avionics (e.g.,
LynxOS-178~\cite{lynxos-178}), automotive (e.g., Tresos, the
operating system defined in AUTOSAR~\cite{autosar}), and
cross-industry domains (DECOS OS~\cite{DECOS}). A comparison of the
mentioned partitioning operating systems can be found in
\cite{PartitioningOS}. They  provide applications shared access to
critical system resources on an integrated computing platform.
Applications may belong to different security domains and can have
different safety-critical impact on the system. To avoid unwanted 
interference between the applications, reliable protection is
guaranteed in both the spatial and the temporal domain that is
achieved by using partitions on the system level. Spatial
partitioning ensures that an application cannot access another 
application's code or data in memory or on disk. Temporal
partitioning guarantees an application access to the critical system
(CPU) resources via dedicated time intervals regardless of other
applications. 


Our approach combines mixed-criticality and partitioning techniques 
to meet the requirements of secure DRE systems.
\iap\ supports multiple levels
of criticality, with tasks being assigned to a single criticality
level. For security and fault isolation reasons, applications are
strictly separated by means of spatial and temporal partitioning, and
applications are required to use a novel secure communication method
for all communications, described in \iap\ \cite{ISIS_F6_Aerospace:12}.

Our work has many similarities with the resource-centric real-time
kernel~\cite{DistributedRK-Rajkumar08} to support real-time
requirements of distributed systems hosting multiple applications. 
Though achieved differently, both frameworks use deployment services
for the automatic deployment of distributed applications, and enforcing 
resource isolation among applications. However, to the best of our knowledge,
\cite{DistributedRK-Rajkumar08} does not include support for process
management, temporal isolation guarantees, partition management, and
secure communication simultaneously. 


\vspace{-0.09in}
\section{DREMS Architecture}
\label{sec:task_model}
\iap\ \cite{ISIS_F6_Aerospace:12,6899124,4813} is a distributed system architecture that consists
of one or more computing nodes grouped into a cluster. It is conceptually similar to the recent Fog Computing Architecture \cite{vaquero2014finding}. Distributed applications, composed from cooperating
processes called \textit{actors}, provide services for the end-user.
Actors specialize the notion of OS processes; 
they have persistent identity that allows them to be transparently
migrated between nodes, and they have strict limits on resources that
they can use.  Each actor is constructed from one or more reusable
components~\cite{ISIS_F6_ISORC:13,4813} where each component is
single-threaded.  
\vspace{-0.05in}

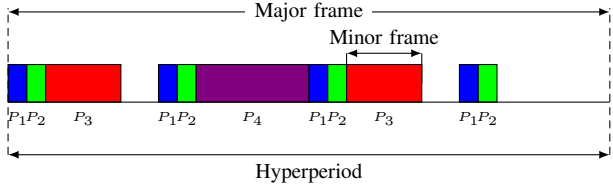
\begin{figure}[t]
\centering
\begin{tikzpicture}[x=1cm,y=1cm, 
every label/.style={font=\tiny},
p1/.style={shape=rectangle,draw,fill=blue, align=center,font=\tiny,minimum height=5mm,minimum width=2.5mm},
p2/.style={shape=rectangle,draw,fill=green,align=center,font=\tiny,minimum height=5mm,minimum width=2.5mm},
p3/.style={shape=rectangle,draw,fill=red,  align=center,font=\tiny,minimum height=5mm,minimum width=10mm},
p4/.style={shape=rectangle,draw,fill=red!50!blue,align=center,font=\tiny,minimum height=5mm,minimum width=15mm},
]
\def\pdist{0.25}\def\hpy{-.95}\def\MF{.95}\def\mfh{.4}\def\mfbeg{17.5*\pdist}\def\mfend{21.5*\pdist}\def\labelpos{270}
\node[p1](P11) at (   0*\pdist,0) [label=\labelpos:$P_1$]{}; \node[p2](P21) at ( 1*\pdist,0) [label=\labelpos:$P_2$] {};
\node[p3](P31) at ( 3.5*\pdist,0) [label=\labelpos:$P_3$]{}; 
\node[p1](P12) at (   8*\pdist,0) [label=\labelpos:$P_1$]{}; \node[p2](P22) at ( 9*\pdist,0) [label=\labelpos:$P_2$] {};
\node[p4](P41) at (12.5*\pdist,0) [label=\labelpos:$P_4$]{}; 
\node[p1](P13) at (  16*\pdist,0) [label=\labelpos:$P_1$]{}; \node[p2](P23) at (17*\pdist,0) [label=\labelpos:$P_2$] {};
\node[p3](P32) at (19.5*\pdist,0) [label=\labelpos:$P_3$]{};
\node[p1](P14) at (  24*\pdist,0) [label=\labelpos:$P_1$]{}; \node[p2](P24) at (25*\pdist,0) [label=\labelpos:$P_2$] {};
\draw[densely dashed,-] (-0.5*\pdist,\hpy) -- (-0.5*\pdist,1); \draw[densely dashed,-] (31.5*\pdist,\hpy) -- (31.5*\pdist,1);
\draw[-] (-0.5*\pdist,-.25) -- (31.5*\pdist,-.25);
\draw[latex-latex] (-0.5*\pdist,\hpy) -- (31.5*\pdist,\hpy) node [below,align=center,midway,font=\footnotesize] {Hyperperiod};
\draw[latex-latex] (-0.5*\pdist,\MF) -- node[sloped,fill=white,font=\footnotesize]{Major frame} (31.5*\pdist,\MF);
\draw[-] (\mfbeg,0) -- (\mfbeg,\mfh); \draw[-] (\mfend,0) -- (\mfend,\mfh);
\draw[latex-latex] (\mfbeg,\mfh) -- (\mfend,\mfh) node [above,align=center,midway,font=\footnotesize] {Minor frame};
\end{tikzpicture}%

\caption{A Major Frame. The four partitions (period,duration) in this frame are $P_1$ (2s, 0.25s), $P_2$ (2s, 0.25s), $P_3$ (4s, 1s), and $P_4$ (8s, 1.5s). }
\label{fig:validschedule}
\vspace{-0.2in}
\end{figure}

\subsection{Partitioning Support}
\label{sec:scheduler}
The system guarantees spatial isolation between actors  by (a)
providing  a separate address space for each actor; (b) enforcing that an
I/O device can be accessed by only one actor at a time; and (c)
facilitating temporal isolation between processes by the scheduler.
Spatial isolation is implemented by the Memory Management Unit of the CPU,  
while temporal isolation is provided via ARINC-653~\cite{ARINC-653} style
\textit{temporal partitions}, implemented in the OS scheduler. 


A temporal partition is characterized by two parameters: period and
duration.  The period reflects how often the tasks of the
partition will be guaranteed CPU allocation.  The duration governs the
length of the CPU allocation window in each cycle.  Given the period
and duration of all temporal partitions, an execution schedule can be
generated by solving a series of constraints, see~\cite {ACM_SPE:10}.
A feasible solution, \emph{e.g.} Figure~\ref{fig:validschedule}, comprises
a repeating frame of windows, where each window is assigned to a
partition.  These windows are called \emph{minor frames}.  The length
of a window assigned to a partition is always the same as the duration
of that partition.  The repeating frame of minor frames, known as the \emph{major frame},
has a length called the \emph{hyperperiod}.  The hyperperiod is the lowest common multiple
of the partition periods.

\subsection{Criticality Levels Supported by the \iap\ OS Scheduler}
\label{sec:criticality_levels}
The \iap\ OS scheduler can manage CPU's time
for tasks on three different criticality levels:
\emph{Critical}, \emph{Application} and \emph{Best Effort}.
The \emph{Critical} tasks provide kernel level services and system
management services. These tasks will be scheduled based on their
priority whenever they are ready. \emph{Application} tasks are
mission specific and are isolated from each other. These tasks are
constrained by temporal partitioning and can be preempted by tasks of the
\emph{Critical} level. Finally, \emph{Best Effort} tasks are executed
whenever no tasks of any higher criticality level are available.

Note that actors in an application can have different criticality levels,
but all tasks associated with an actor must have the same criticality
level, \emph{i.e.} an actor cannot have both \emph{Critical} tasks and
\emph{Application} tasks.

\subsection{Multiple partitions}
\label{sec:datastructure}
To support the different levels of criticality, we extend the \textit{runqueue} data structure of the Linux kernel \cite{garg2009real}. A runqueue maintains a list of tasks eligible for scheduling. 
In a multicore system, this structure is replicated per CPU. In a fully preemptive mode, the scheduling decision is made by  evaluating which task should be executed next on a CPU when an interrupt handler exits, when a system call returns, or when the scheduler function is explicitly invoked to preempt the current process.
We created one runqueue per temporal partition per CPU. Currently, the system can support 64 {\it Temporal partitions}, also referred to as Application partitions in the sequel. One extra runqueue is created for the critical tasks. These tasks are said to belong to the {\it System partition}.  The Best effort tasks are managed through the Linux Completely Fair Scheduler (default) runqueue and are considered for execution as part of the System partition when no other tasks are eligible to run.

\subsection{CPU Cap and Work Conserving Behavior}
\label{sec:CPUCAP}
The schedulability of the \emph{Application} level tasks is
constrained by the current load coming from the \emph{Critical} tasks
and the temporal partitioning used on the \emph{Application} level.
Should the load of the \emph{Critical} tasks exceed a threshold the system
will not be able to schedule tasks on the \emph{Application} level. A
formal analysis of the response-time of the \emph{Application} level
tasks will not be provided in this paper, however, we present a 
description of the method we will use to address the analysis which
will build on available results from
\cite{BaruahRTA4MCS, PartitionedRTA-Almeida04, 
HierarchicalRTA-Lipari05}.

The submitted load function $H_i(t)$ determines the maximum load
submitted to a partition by the task $\tau_i$ itself after its
release together with all higher priority tasks belonging to
the same partition. The availability function $A_S(t)$ returns for
each time instant the cumulative computation time available for the
partition to execute tasks. In the original model~\cite{PartitionedRTA-Almeida04}
$A_S(t)$ is the availability function of a periodic server.
 The response-time of a task $\tau_i$ is the time
when $H_i(t)$ intersects the availability function $A_S(t)$ for the
first time.  In our system $A_S(t)$ is decreased by the load of the
available \emph{Critical} tasks which, if unbounded, could block the
application level tasks forever. This motivates us to enforce a bound
on the load of the \emph{Critical} tasks. This bound is referred to as
{\bf CPU cap}.

In \iap\ OS, the CPU cap can be applied to tasks on the
\emph{Critical} and \emph{Application} level to provide scheduling
fairness within a partition or hyperperiod. Between criticality
levels, the CPU cap provides the ability to prevent higher
criticality tasks from starving lower criticality tasks of the CPU.
On the \emph{Application} level, the CPU cap can be used to bound
the CPU consumption of higher priority tasks to allow the execution
of lower priority tasks inside the same partition. If the CPU cap
enforcement is enabled, then it is possible to set a maximum CPU
time that a task can use, measured over a configurable number of
major frame cycles.

The CPU cap is enforced in a work conserving manner, \textit{i.e.}, 
if a task has reached its CPU cap but there are no other
available tasks, the scheduler will continue scheduling the task past
its ceiling. In case of \emph{Critical} tasks, when the CPU cap is reached,
the task is not marked ready for execution unless
(a) there is no other ready task in the system; or 
(b) the CPU cap accounting is reset.
This behavior ensures that the kernel tasks, such as those belonging
to network communication, do not
overload the system, for example in a denial-of-service attack.
For the tasks on the \emph{Application} level, the CPU cap is
specified as a percentage of the total duration of the partition,
the number of major frames, and the number of CPU cores
available all multiplied together. When an \emph{Application} task reaches
the CPU cap, it is not eligible to be scheduled again unless 
the following is true: either
(a) there are no \emph{Critical} tasks to schedule and there are no other ready tasks in the partition; or (b) the CPU cap accounting has been reset.

\subsection{Dynamic Major Frame Configuration}
\label{sec:reconfiguration}

During the configuration process that can be repeated at
any time without rebooting the node, the kernel receives the major
frame structure that contains a list of minor frames and it also
contains the length of the hyperperiod, partition periodicity, and
duration. Note that major frame reconfiguration can only be
performed by an actor with suitable capabilities.  More details on the
\iap\ capability model can be found in~\cite{ISIS_F6_Aerospace:12}.

Before the frames are set up, the process configuring the frame has to
ensure that the following three constraints are met: (C0) The
hyperperiod must be the least common multiple of partition periods;
(C1) The offset between the major frame start and the first minor
frame of a partition must be less than or equal to the partition
period:  $(\forall p \in \mathbb{P})(O_{1}^{p} \leq \phi(p))$; (C2)
Time between any two executions should be equal to the partition
period: $(\forall p \in
\mathbb{P})(k\in[1,N(p)-1])(O_{k+1}^{p}=O_{k}^{p}+ \phi(p))$, where
$\mathbb{P}$ is the set of all partitions, $N(p)$ is the number of
partitions, $\phi(p)$ is the period of partition $p$ and $\Delta(p)$
is the duration of the partition $p$. $O^p_i$ is the offset of
$i^{th}$ minor frame for partition $p$ from the start of the major
frame, $H$ is the hyperperiod.

The kernel checks two additional constraints: (1) All minor frames
finish before the end of the hyperperiod: $(\forall i)(O_{i}.start+O_{i}.duration
\leq H)$ and (2) minor frames cannot overlap, i.e. given a sorted minor
frame list (based on their offsets): $(\forall i <
N(O))(O_{i}.start+O_{i}.duration \leq O_{i+1})$, where $N(O)$ is the number
of minor frames.   Note that the minor frames need not be contiguous,
as the update procedure fills in any gaps automatically.

\begin{figure}[t]
\centering
\includegraphics[width=0.48\textwidth]{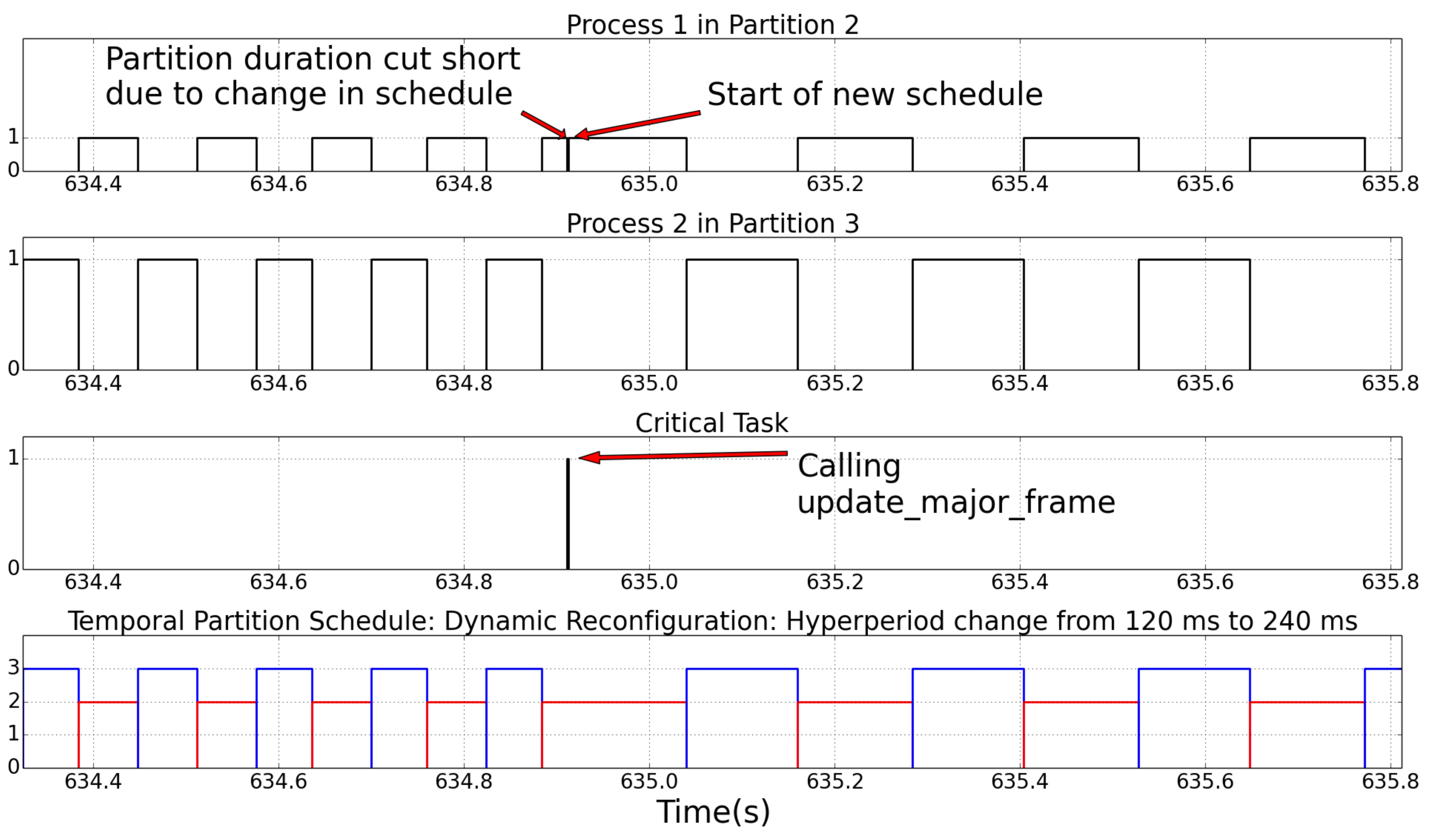}
\caption{Two single-threaded processes run in separate partitions with a duration of $60 ms$ each. The schedule is dynamically reconfigured so that each partition duration is doubled. 
A \emph{Critical} task is responsible for calling the update\_major\_frame system call. Duration of the active partition is cut short at the point when update\_major\_frame function is called. 
}
\label{fig:dynamic_reconfig}
\end{figure}

If the constraints are satisfied, then the task is moved to the first core, \emph{CPU0} if it is not already on \emph{CPU0}. 
This is done because the global tick (explained in next subsection) used for implementing the major 
frame schedule is also executed on \emph{CPU0}. By moving the task to \emph{CPU0} and disabling interrupts, 
the scheduler ensures that the current frame is not changed while the major frame is being updated. 
At this point the task also obtains a spin lock to  ensure that no other task can update the major frame at 
the same time. In this procedure the scheduler state is also set to \texttt{APP\_INACTIVE} (see Table \ref{table:variable}), to stop the scheduling of all application
tasks across other cores. The main scheduling loop reads the scheduler state before scheduling application tasks. A  scenario showing dynamic reconfiguration can be seen in Figure~\ref{fig:dynamic_reconfig}. 

\begin{table}[ht]
\centering
\caption{The states of the DREMS Scheduler}
\footnotesize
\begin{tabular}{| c | p{0.3\textwidth} |}
\hline
 APP\_INACTIVE &Tasks in temporal partitions are not run \\\hline
 APP\_ACTIVE &Inverse of APP\_INACTIVE\\\hline
\end{tabular}
\label{table:variable}
\end{table}

It is also possible to set the global tick (that counts the hyperperiods) to be started with an offset. 
This delay can be used to synchronize the start of the hyperperiods across nodes of the cluster. This is necessary to 
ensure that all nodes schedule related temporal partitions at the same time.
This ensures that for an application that is distributed
across multiple nodes, its \emph{Application} level tasks run at
approximately the same time on all the nodes which enables low latency
communication between dependent tasks across the node level. 


%
	
\vspace{-0.1in}
\subsection{Main Scheduling Loop}	
\label{sec:scheduling}

 A periodic tick running at $250$ Hz\footnote{The kernel tick value is also called 'jiffy' and can be set to a different value when the kernel image is compiled}  is used to ensure
that a scheduling decision is triggered at least every $4$ ms.  This
tick runs with the base clock of \emph{CPU0} and executes a procedure called $GlobalTick$
in the interrupt context only on
\emph{CPU0}.   
 This procedure enforces the partition scheduling and
updates the current minor frame and hyperperiod start time
(\texttt{HP\_start}).  The partition schedule is determined by
a circular linked list of minor frames which comprise
the major frame.  Each entry in this list contains that partition's duration,
so the scheduler can easily calculate when to switch to the next minor frame. 

After the global tick handles the partition switching, the function to
get the next  runnable task is invoked. This function combines the
\emph{mixed criticality} scheduling with the \emph{temporal partition}
scheduling. For mixed
criticality scheduling, the \emph{Critical} system tasks should preempt
the \emph{Application} tasks, which themselves should preempt the
\emph{Best Effort} tasks. This policy is implemented by  \emph{Pick\_Next\_Task} subroutine, which is called first for the system partition.
Only if there are no runnable \emph{Critical} system tasks and the
scheduler state is not inactive, i.e. the application partitions are allowed to run\footnote{The OS provides support for pausing all application partitions and ensuring that only system partition is executed}, will
\emph{Pick\_Next\_Task} be called for the \emph{Application} tasks.
Thus, the scheduler does not schedule any \emph{Application} tasks during
a major frame reconfiguration. Similarly \emph{Pick\_Next\_Task} will
only be called for the \emph{Best Effort} tasks if there are both no
runnable \emph{Critical} tasks and no runnable \emph{Application} tasks.

\subsection{Pick\_Next\_Task and CPU Cap}
The \emph{Pick\_Next\_Task} function returns  either the highest
priority task from the current temporal partition (or the system
partition, as application) or an empty list of there are no runnable
tasks.  
 If CPU cap is disabled, the
\emph{Pick\_Next\_Task} algorithm returns the first task from the specified
runqueue. For the best effort class, the default algorithm for the
Completely Fair Scheduler policy in the Linux Kernel
\cite{mauerer2008} is used.

  If the CPU cap is enabled,
the \emph{Pick\_Next\_Task} algorithm iterates through the task list
at the highest priority index of the runqueue, because unlike the
Linux scheduler, the tasks may have had their disabled bit set by the
scheduler if it had enforced their CPU cap.  If the algorithm finds a
disabled task in the task list, it checks to see when it was disabled;
if the task was disabled in the previous CPU cap window, it reenables the
task and sets it as the $next\_task$.  If, however, the task
was disabled in the current CPU cap window, the algorithm continues
iterating through the task list until it finds a task which is
enabled.  If the algorithm finds no enabled task, it returns the first
task from the list if the current runqueue belongs to an application partition. 


This iteration through the task list when CPU cap
enforcement is enabled increases the complexity of the scheduling algorithm to
$O(n)$, where $n$ is the number of tasks in that temporal partition,
compared to the Linux scheduler's complexity of $O(1)$.  Note that
this complexity is incurred when CPU cap enforcement is
enabled and there is at least one actor that has partial CPU cap (less
than 100\%).  In the worst case, if all actors are given a partial CPU
cap, the scheduler performance may degrade necessitating more
efficient data structures.

\begin{figure}[t]
\centering
\includegraphics[width=0.5\textwidth]{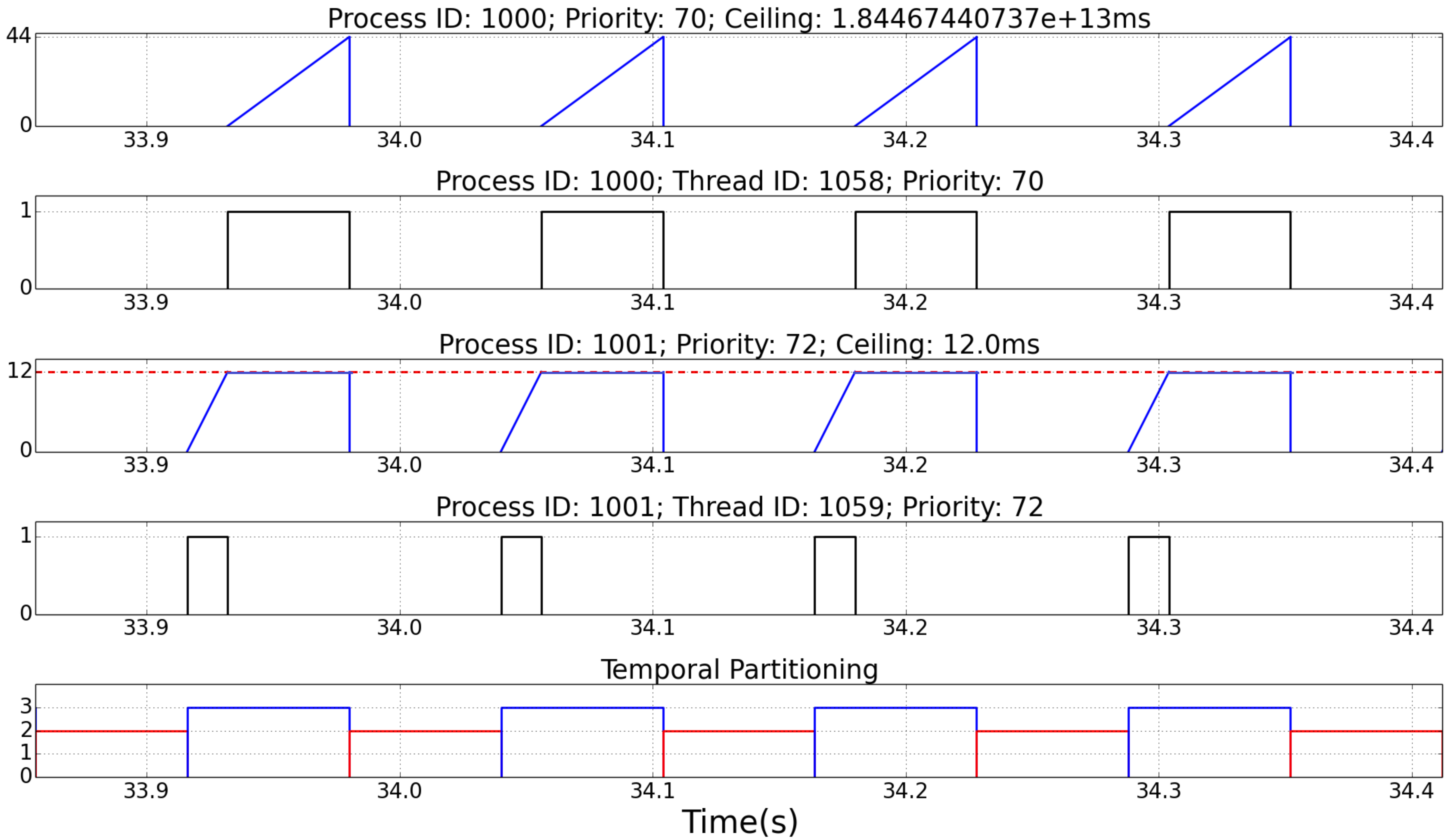}
\caption{Single Threaded processes 1000 and 1001 share a partition with a duration of $60 ms$.
Process 1000 has 100\% CPU cap and priority 70; process 1001 has 20\% CPU cap, and higher priority 72.
Since process 1001 has a CPU cap less than 100\%, a ceiling is calculated for this process: $20\%$ of $60 ms$ = $12 ms$. The average jitter was calculated to be 2.136 ms with a maximum jitter of 4.0001 ms. 
}
\label{fig:scenario1}
\vspace{-0.2in}
\end{figure}

To complete the enforcement of the CPU cap, the scheduler updates the
statistics tracked about the task and then updates the disabled bit of
the task accordingly.
Figure~\ref{fig:scenario1}, shows
the above mentioned scheduler decisions when CPU cap is placed on
processes that share a temporal partition.  
To facilitate analysis, the scheduler uses a logging framework that
updates a log every time a context switch happens.  Figure~\ref{fig:scenario1} clearly shows the lower priority actor
executing after the higher priority actor has reached its CPU cap. 

\begin{figure*}[!htb]
\begin{subfigure}[b]{0.9\textwidth}
\centering
\includegraphics[width=0.9\textwidth]{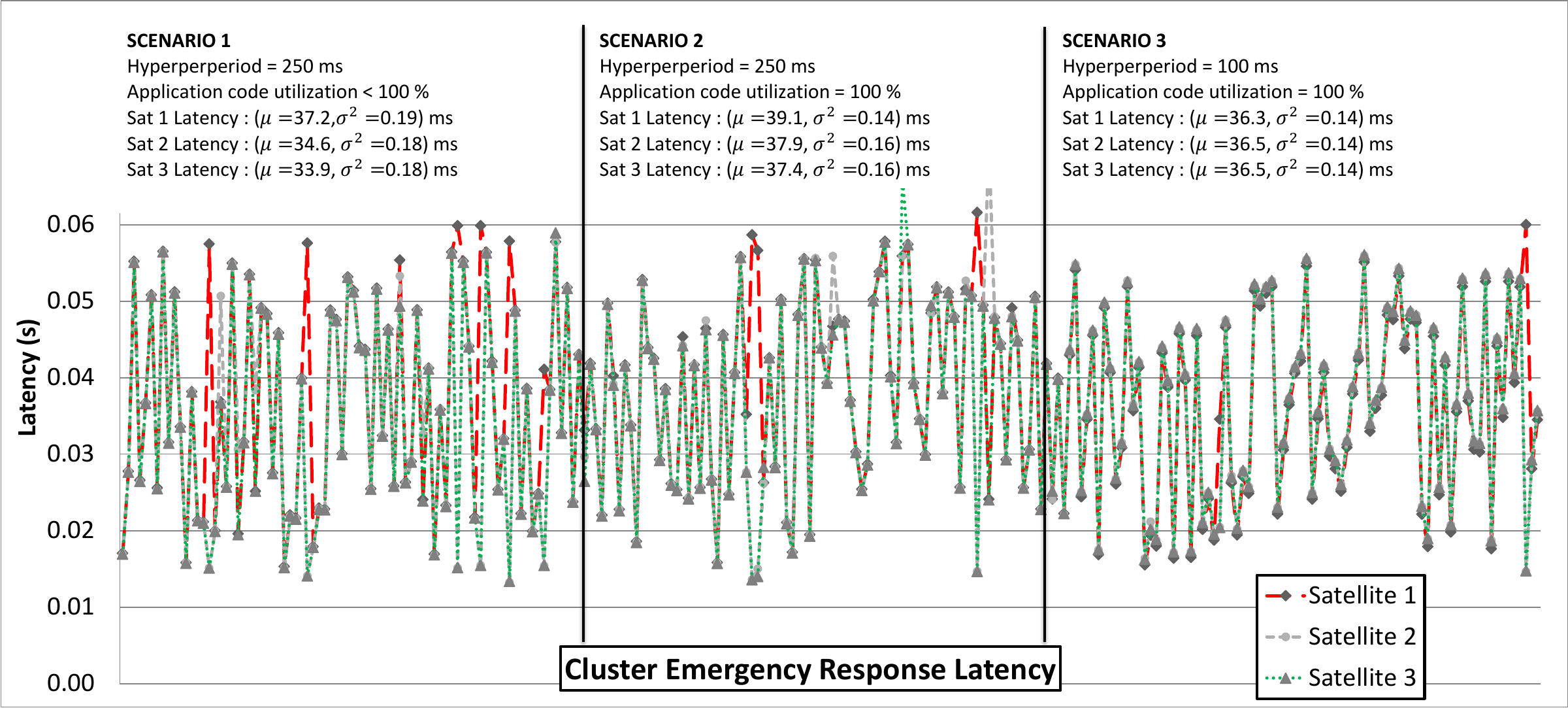}
\caption{
This is the time between reception of the \emph{scatter} command by satellite 1 and the activation of the thrusters on each satellite, corresponding to interactions \textit{CommandProxy} to \textit{ModuleProxy}. The three regions of the plot indicate the three scenarios: (1) image processing application has limited use of its partitions and has a hyperperiod of 250 $ms$, (2) image processing application has full use of its partitions and has a hyperperiod of 250 $ms$, and (3) image processing application has full use of its partitions and has a hyperperiod of 100 $ms$.  The averages and variances for the satellites' latencies are shown for each of the three scenarios.}
\label{fig:latency_analysis}
\end{subfigure}
\begin{subfigure}[b]{0.9\textwidth}
\centering
\includegraphics[width=0.9\textwidth]{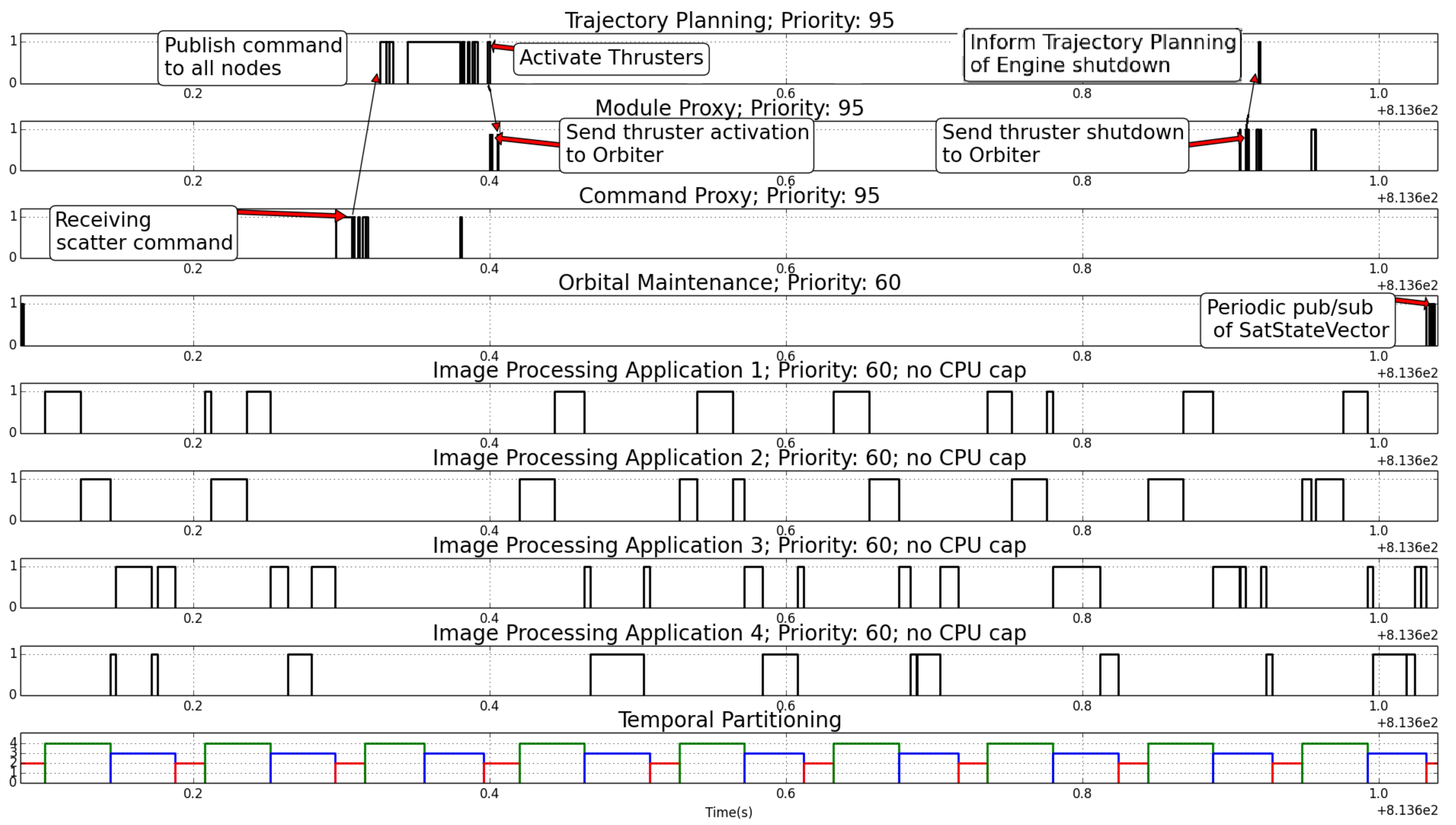}
\caption{The engine activation following reception of a \emph{scatter} command is annotated for the relevant actors for scenario 2 shown above. The \emph{scatter} command causes the \emph{TrajectoryPlanning} to request \emph{ModuleProxy} to activate the thrusters for 500 $ms$. Notice that the image processing does not run while the mission-critical tasks are executing - without halting the partition scheduling.  Also note that the context switching during the execution of the critical tasks is the execution of the secure transport kernel thread. Only the application tasks are shown in the log; the kernel threads and other background processes are left out for clarity.}
\label{fig:orbiter_plot}
\end{subfigure}
\vspace{-0.1in}
\caption{\iap\ Mixed Criticality Demo }
\vspace{-0.2in}
\end{figure*}

\section{Experiment: A 3-node Satellite Cluster}
\label{sec:experiment}

To demonstrate the DREMS platform, a multi-computing node experiment was created on a cluster of fanless computing nodes with a 1.6 GHz Intel Atom N270 processor and 1 GB of RAM each. 
On these nodes, a cluster of three satellites was emulated and each satellite ran the example applications described in Section~\ref{sec:intro}. 
  Because the performance of the cluster flight control application is of interest, we explain the interactions between its actors below.  

\begin{figure}[t]
\centering
\begin{tikzpicture}[x=1cm,y=1cm, 
taskW/.style={shape=circle,draw,align=center,font=\tiny,text=white},
taskB/.style={shape=circle,draw,align=center,font=\tiny}]
\def\uxo{0}\def\uyo{3}\def\lxo{0}\def\lyo{0}
\node[taskB](M11) at (\uxo-2,\uyo+1) {$M^1_1$}; \node[taskB](V11) at (\uxo,\uyo+1) {$O^1_1$}; \node[taskB](V21) at (\uxo+2,\uyo+1) {$O^2_1$};
\node[taskB](M12) at (\uxo-2,\uyo) {$M^1_2$};   \node[taskB](V12) at (\uxo,\uyo) {$O^1_2$};   \node[taskB](V22) at (\uxo+2,\uyo) {$O^2_2$};
\node[taskB](M13) at (\uxo-2,\uyo-1) {$M^1_3$}; \node[taskB](V13) at (\uxo,\uyo-1) {$O^1_3$}; \node[taskB](V23) at (\uxo+2,\uyo-1) {$O^2_3$};
\draw[-latex] (M11.0) -- (V11.180); \draw[-latex] (M12.0) -- (V12.180); \draw[-latex] (M13.0) -- (V13.180);
\draw[-latex] (V11.0) -- (V21.180); \draw[-latex] (V12.0) -- (V21.200); \draw[-latex] (V13.0) -- (V21.220);
\draw[-latex] (V11.0) -- (V22.160); \draw[-latex] (V12.0) -- (V22.180); \draw[-latex] (V13.0) -- (V22.200);
\draw[-latex] (V11.0) -- (V23.140); \draw[-latex] (V12.0) -- (V23.160); \draw[-latex] (V13.0) -- (V23.180);
\node[taskB](C11) at (\lxo-3,\lyo+1) {$C^1_1$}; \node[taskB](O11) at (\lxo-1,\lyo+1) {$T^1_1$}; \node[taskB](O21) at (\lxo+1,\lyo+1) {$T^2_1$}; \node[taskB](M21) at (\lxo+3,\lyo+1) {$M^2_1$};
\node[taskB](C12) at (\lxo-3,\lyo)   {$C^1_2$}; \node[taskB](O12) at (\lxo-1,\lyo)   {$T^1_2$}; \node[taskB](O22) at (\lxo+1,\lyo)   {$T^2_2$}; \node[taskB](M22) at (\lxo+3,\lyo) {$M^2_2$};
\node[taskB](C13) at (\lxo-3,\lyo-1) {$C^1_3$}; \node[taskB](O13) at (\lxo-1,\lyo-1) {$T^1_3$}; \node[taskB](O23) at (\lxo+1,\lyo-1) {$T^2_3$}; \node[taskB](M23) at (\lxo+3,\lyo-1) {$M^2_3$};
\draw[-latex,very thick] (C11.0) -- (O11.180); \draw[-latex] (C12.0) -- (O12.180); \draw[-latex] (C13.0) -- (O13.180);
\draw[-latex,very thick] (O11.0) -- (O21.180); \draw[-latex] (O12.0) -- (O21.200); \draw[-latex] (O13.0) -- (O21.220);
\draw[-latex,very thick] (O11.0) -- (O22.160); \draw[-latex] (O12.0) -- (O22.180); \draw[-latex] (O13.0) -- (O22.200);
\draw[-latex,very thick] (O11.0) -- (O23.140); \draw[-latex] (O12.0) -- (O23.160); \draw[-latex] (O13.0) -- (O23.180);
\draw[-latex,very thick] (O21.0) -- (M21.180); \draw[-latex,very thick] (O22.0) -- (M22.180); \draw[-latex,very thick] (O23.0) -- (M23.180);
\end{tikzpicture}%
\vspace{1.5mm}

\begin{footnotesize}
\begin{tabular}{| c | c | c | p{4cm} |}
\hline
Task & Actor & Activity \\ [0.5ex]
\hline
$M^1$ & \multirow{2}{*}{$ModuleProxy$} &  Inform $O^1$ of new state\\
\cline{1-1}\cline{3-3}
$M^2$ & & Activate engine\\
\hline
$O^1$ & \multirow{2}{*}{$OrbitalMaintenance$} & Publish new state\\
\cline{1-1}\cline{3-3}
$O^2$ & & Subscribe to new state\\
\hline
$T^1$ & \multirow{2}{*}{$TrajectoryPlanning$} & Publish new command\\
\cline{1-1}\cline{3-3}
$T^2$ & & Subscribe to new command\\
\hline
$C^1$ & $CommandProxy$ & Inform $T^1$ of command\\
\hline
\end{tabular}
\end{footnotesize}

\caption{\iap\ tasks : 
\emph{ModuleProxy} tasks control thruster activation in Orbiter and state vector retrieval from Orbiter.
\emph{OrbitalMantenance} tasks track the cluster satellites' state vectors and disseminate them.  
\emph{TrajectoryPlanning} tasks control the response to commands and satellite thruster activation.
\emph{CommandProxy} tasks inform the satellite of a command from the ground network.  
For these tasks, the subscript represents the node ID on which the task is deployed.
The total latency of the interaction $C_1^1 \to M_N^2$ represents the total emergency response latency
between receiving the \emph{scatter} command and activating the thrusters.  This interaction pathway 
is shown in bold.
}
\label{fig:OrbiterDemoTDG}
\vspace{-0.2in}
\end{figure}
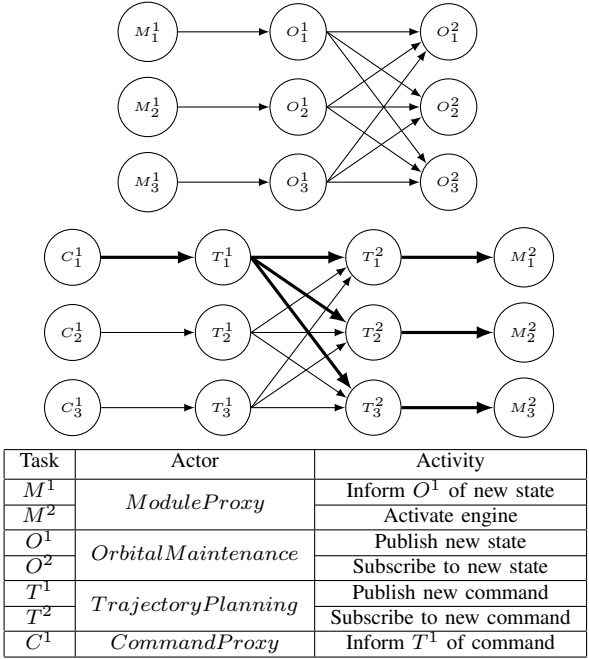

The mission-critical cluster flight application (CFA) (Figure \ref{fig:OrbiterDemoTDG}) consists of four actors: \emph{OrbitalMaintenance}, \emph{TrajectoryPlanning}, \emph{CommandProxy}, and \emph{ModuleProxy}. \emph{ModuleProxy} connects to the Orbiter space
flight simulator (\url{http://orbit.medphys.ucl.ac.uk/}) that simulates the satellite hardware and orbital mechanics for the three satellites in low Earth orbit. \emph{CommandProxy} receives commands from the ground network.  \emph{OrbitalMaintenance} keeps track of every satellite's position and updates the cluster with its current position. This is done by a group publish subcribe interaction between all \textit{OrbitalMaintenance} actors across all nodes.  

 Additionally, four image processing application (IPA) actors (one
 actor per application instance) are deployed as application
 tasks. The IPA design allows the percentage of CPU cycles consumed by
 them to be configurable.  The four IPAs are  
 assigned to two partitions, such that each partition contains two IPA
 actors.  A third, shorter, partition runs the
 \emph{OrbitalMaintenance} actor; since it is a periodic task, it
 updates the satellite state every second and is not critical in an
 emergency.   

Figures~\ref{fig:latency_analysis} and~\ref{fig:orbiter_plot} show the results from three different scenarios: 1) hyperperiod of 250 ms, with IPA consuming less than 50 percent CPU. 2)  hyperperiod of 250 ms, with IPA consuming 100 percent CPU and 3)  hyperperiod of 100 ms, with IPA consuming 100 percent CPU. As shown in figure ~\ref{fig:latency_analysis}, the emergency response latency over the three nodes was quite low with very little variance, and did not correlate with either the image application's CPU utilization or the application's partition schedule.  Since we show that the emergency response has very low latency with little variance between different application loads on the system, we provide a stable platform for deterministic and reliable emergency response.  As such, the satellite cluster running the \iap\ infrastructure is able to quickly respond to emergency situations despite high application CPU load and without altering the partition scheduling.  Figure~\ref{fig:orbiter_plot}  demonstrates the proper preemption of the image processing tasks by the critical CFA tasks for scenario 2.

\vspace{-0.1in}
\section{Conclusions and Future Work}
\label{sec:conclusions}

This paper propounds the notion of managed distributed real-time and
embedded (DRE) systems that are deployed in mobile computing
environments. 
To that end, we described 
the design and implementation of a
distributed operating system called \iap\ OS focusing on a key
mechanism: the scheduler. 
We have verified the behavioral properties of the OS
scheduler, focusing on temporal and spatial process isolation, safe
operation with mixed criticality, precise control of process CPU
utilization and dynamic partition schedule reconfiguration.  We have
also analyzed the scheduler  properties of a
distributed application built entirely using this platform and hosted
on an emulated cluster of satellites.

We  are extending this operating system  to build an open-source FACE\textsuperscript{tm} Operating System Segment \cite{face}, called COSMOS (\textbf{C}ommon \textbf{O}perating \textbf{S}ystem for \textbf{M}odular \textbf{O}pen \textbf{S}ystems). To the best of our knowledge this is the first open source implementation of its kind that provides both ARINC-653 and POSIX partitions.


\textbf{Acknowledgments:} This work was supported by the DARPA 
under contract NNA11AC08C. Any opinions, findings, and
conclusions or recommendations expressed in this material
are those of the author(s) and do not  reflect
the views of DARPA.
\balance

\bibliographystyle{IEEEtran}
\bibliography{f6}

\begin{thebibliography}{10}
\providecommand{\url}[1]{#1}
\csname url@samestyle\endcsname
\providecommand{\newblock}{\relax}
\providecommand{\bibinfo}[2]{#2}
\providecommand{\BIBentrySTDinterwordspacing}{\spaceskip=0pt\relax}
\providecommand{\BIBentryALTinterwordstretchfactor}{4}
\providecommand{\BIBentryALTinterwordspacing}{\spaceskip=\fontdimen2\font plus
\BIBentryALTinterwordstretchfactor\fontdimen3\font minus
  \fontdimen4\font\relax}
\providecommand{\BIBforeignlanguage}[2]{{%
\expandafter\ifx\csname l@#1\endcsname\relax
\typeout{** WARNING: IEEEtran.bst: No hyphenation pattern has been}%
\typeout{** loaded for the language `#1'. Using the pattern for}%
\typeout{** the default language instead.}%
\else
\language=\csname l@#1\endcsname
\fi
#2}}
\providecommand{\BIBdecl}{\relax}
\BIBdecl

\bibitem{ARINC-653}
\emph{{Document No. 653: Avionics Application Software Standard Inteface (Draft
  15)}}, {ARINC Incorporated}, Annapolis, Maryland, USA, Jan. 1997.

\bibitem{6899124}
G.~Karsai, D.~Balasubramanian, A.~Dubey, and W.~R. Otte, ``Distributed and
  managed: Research challenges and opportunities of the next generation
  cyber-physical systems,'' in \emph{2014 IEEE 17th International Symposium on
  Object/Component/Service-Oriented Real-Time Distributed Computing}, June
  2014, pp. 1--8.

\bibitem{Vestal2007}
S.~Vestal, ``{Preemptive Scheduling of Multi-Criticality Systems with Varying
  Degrees of Execution Time Assurance},'' in \emph{Proc.\ of 28th IEEE
  Real-Time Systems Symposium}, Tucson, AZ, Dec. 2007, pp. 239--243.

\bibitem{BaruahRTA4MCS}
S.~Baruah, A.~Burns, and R.~Davis, ``{Response-Time Analysis for
  Mixed-Criticality Systems},'' in \emph{Proceedings of the 2011 32nd IEEE
  Real-Time Systems Symposium}, Vienna, Austria, Nov. 2011, pp. 34--43.

\bibitem{lynxos-178}
\BIBentryALTinterwordspacing
LynuxWorks, ``{RTOS for Software Certification: LynxOS-178}.'' [Online].
  Available: \url{http://www.lynuxworks.com/rtos/rtos-178.php}
\BIBentrySTDinterwordspacing

\bibitem{autosar}
\BIBentryALTinterwordspacing
{Autosar GbR}, ``{AUTomotive Open System ARchitecture},''
  \url{http://www.autosar.org/}. [Online]. Available:
  \url{http://www.autosar.org/}
\BIBentrySTDinterwordspacing

\bibitem{DECOS}
R.~Obermaisser, P.~Peti, B.~Huber, and C.~E. Salloum, ``{DECOS: An Integrated
  Time-Triggered Architecture},'' \emph{e\&i journal (Journal of the Austrian
  Professional Institution for Electrical and Information Engineering)}, vol.
  123, no.~3, pp. 83--95, Mar. 2006.

\bibitem{PartitioningOS}
B.~Leiner, M.~Schlager, R.~Obermaisser, and B.~Huber, ``{A Comparison of
  Partitioning Operating Systems for Integrated Systems},'' in \emph{Computer
  Safety, Reliability and Security}, ser. Lecture Notes in Computer
  Science.\hskip 1em plus 0.5em minus 0.4em\relax Springer, 2007, vol.
  4680/2007, pp. 342--355.

\bibitem{ISIS_F6_Aerospace:12}
A.~Dubey, W.~Emfinger, A.~Gokhale, G.~Karsai, W.~Otte, J.~Parsons, C.~Szabo,
  A.~Coglio, E.~Smith, and P.~Bose, ``{A Software Platform for Fractionated
  Spacecraft},'' in \emph{Proceedings of the IEEE Aerospace Conference,
  2012}.\hskip 1em plus 0.5em minus 0.4em\relax Big Sky, MT, USA: IEEE, Mar.
  2012, pp. 1--20.

\bibitem{DistributedRK-Rajkumar08}
K.~Lakshmanan and R.~Rajkumar, ``{Distributed Resource Kernels: OS Suppport for
  End-To-End Resource Isolation },'' in \emph{Proceedings of the 2008 IEEE
  Real-Time and Embedded Technology and Applications Symposium}, St.~Louis, MO,
  Apr. 2008, pp. 195--204.

\bibitem{4813}
S.~Eisele, I.~Madari, A.~Dubey, and G.~Karsai, ``Riaps:resilient information
  architecture platform for decentralized smart systems,'' in \emph{20th IEEE
  International Symposium On Real-Time Computing}, IEEE.\hskip 1em plus 0.5em
  minus 0.4em\relax Toronto, Canada: IEEE, 05/2017 2017.

\bibitem{vaquero2014finding}
L.~M. Vaquero and L.~Rodero-Merino, ``Finding your way in the fog: Towards a
  comprehensive definition of fog computing,'' \emph{ACM SIGCOMM Computer
  Communication Review}, vol.~44, no.~5, pp. 27--32, 2014.

\bibitem{ISIS_F6_ISORC:13}
W.~R. Otte, A.~Dubey, S.~Pradhan, P.~Patil, A.~Gokhale, G.~Karsai, and
  J.~Willemsen, ``{F6COM: A Component Model for Resource-Constrained and
  Dynamic Space-Based Computing Environment},'' in \emph{Proceedings of the
  16th IEEE International Symposium on Object-oriented Real-time Distributed
  Computing (ISORC '13)}, Paderborn, Germany, Jun. 2013.

\bibitem{ACM_SPE:10}
A.~Dubey, G.~Karsai, and N.~Mahadevan, ``{A Component Model for Hard Real-time
  Systems: \textsc{CCM} with \textsc{ARINC-653}},'' \emph{Software: Practice
  and Experience}, vol.~41, no.~12, pp. 1517--1550, 2011.

\bibitem{garg2009real}
A.~Garg, ``Real-time linux kernel scheduler,'' \emph{Linux Journal}, vol. 2009,
  no. 184, p.~2, 2009.

\bibitem{PartitionedRTA-Almeida04}
L.~Almeida and P.~Pedreiras, ``{Scheduling within Temporal Partitions:
  Response-time Analysis and Server Design},'' in \emph{Proc.\ of the 4th ACM
  Int Conf on Embedded Software}, Pisa, Italy, Sep. 2004, pp. 95--103.

\bibitem{HierarchicalRTA-Lipari05}
G.~Lipari and E.~Bini, ``{A Methodology for Designing Hierarchical Scheduling
  Systems},'' \emph{Journal of Embedded Computing}, vol.~1, no.~2, pp.
  257--269, Apr. 2005.

\bibitem{mauerer2008}
\BIBentryALTinterwordspacing
W.~Mauerer, \emph{Professional Linux Kernel Architecture}, ser. Wrox
  professional guides.\hskip 1em plus 0.5em minus 0.4em\relax Wiley, 2008.
  [Online]. Available: \url{http://books.google.com/books?id=4eCr9dr0uaYC}
\BIBentrySTDinterwordspacing

\bibitem{face}
\BIBentryALTinterwordspacing
OpenGroup. [Online]. Available: \url{http://www.opengroup.org/face}
\BIBentrySTDinterwordspacing

\end{thebibliography}

\end{document}